\documentclass[a4paper,fleqn,usenatbib]{mnras}

\usepackage{newtxtext,newtxmath}

\usepackage[T1]{fontenc}
\usepackage{ae,aecompl}

\usepackage[pdftex]{graphicx}	
\usepackage{epstopdf}
\usepackage{amsmath}	
\usepackage{amssymb}	
\usepackage{xspace}
\usepackage{textcomp}
\usepackage{gensymb}
\usepackage{enumerate}
\usepackage{hyperref}

\usepackage{ifthen}
\def\draftversion{0} 

\makeatletter
\newcommand\mytoc{%
    \@starttoc{toc}%
}
\makeatother

\ifthenelse{\equal{\draftversion}{0}}{
	\usepackage{xcolor}
	\newcommand{\tmp}{}
	\newenvironment{envcomm}[1]{\renewcommand{\tmp}{#1}\begin{color}{blue}\begin{center}\hrule\vspace{0.5mm}\tmp's COMMENTS\end{center}}{\begin{center}END OF \tmp's COMMENTS\vspace{0.5mm}\hrule\end{center}\end{color}}
	\newenvironment{draft}{\begin{color}[rgb]{0,0.4,0}\begin{center}\hrule\vspace{0.5mm}DRAFT\end{center}}{\begin{center}END OF DRAFT\vspace{0.5mm}\hrule\end{center}\end{color}}
	\newcommand{\comcomm}[2]{\begin{color}{blue}\ $\bullet$ \textbf{#1:} #2 $\bullet$\ \end{color}}
	\newcommand{\revend}[1]{\par\begin{color}[rgb]{0,0.4,0}\begin{center}\hrule\vspace{0.5mm}END OF #1's REVISIONS\vspace{0.5mm}\hrule\end{center}\end{color}\par}
	\newcommand{\todo}[1]{\begin{color}{red}\ $\bullet$ \textbf{To do: }#1 $\bullet$\ \end{color}}

	\newcommand{\del}[1]{\begin{color}[rgb]{0,0.5,0.0}\ $\bullet$ \textbf{Deleted: }#1 $\bullet$\ \end{color}}
	\newcommand{\sk}[1]{\begin{color}[rgb]{0.6,0,0.6}#1\end{color}}
	\newcommand{\toc}{\par\begin{color}[rgb]{0.6,0,0.6}\begin{center}\hrule\vspace{0.5mm}\begingroup\small\let\cleardoublepage\relax\let\clearpage\relax\mytoc\endgroup\vspace{0.5mm}\hrule\end{center}\end{color}\par}
	}{
	\newsavebox{\trashcan}
	\newenvironment{envcomm}[1]{\begin{lrbox}{\trashcan}\begin{minipage}{\columnwidth}}{\end{minipage}\end{lrbox}}
	
	\newcommand{\comcomm}[2]{}
	\newcommand{\revend}[1]{}
	\newcommand{\todo}[1]{}
	
	\newcommand{\del}[1]{}
	\newcommand{\sk}[1]{}
	\newcommand{\toc}{}
	}

\long\def\symbolfootnote[#1]#2{\begingroup%
\def\thefootnote{\fnsymbol{footnote}}\footnote[#1]{#2}\endgroup} 

\newcommand{\bb}[1]{\ifmmode \mbox{\boldmath $ #1$} \else  \mbox{\boldmath $#1$} \fi}

\newcommand{\U}[1]{\ensuremath{\mathrm{~#1}}}
\newcommand{\yr}{\U{yr}}
\newcommand{\Myr}{\U{Myr}}
\newcommand{\Gyr}{\U{Gyr}}
\newcommand{\pc}{\U{pc}}
\newcommand{\kpc}{\U{kpc}}

\newcommand{\msun}{\U{M}_{\odot}}
\newcommand{\Msun}{\msun}
\newcommand{\Msunyr}{\Msun\yr^{-1}}
\newcommand{\cc}{\U{cm^{-3}}}
\newcommand{\mpc}{\U{M_{\odot}\ pc^{-3}}}

\newcommand{\ramses}{\texttt{RAMSES}\xspace}

\title[star clusters and feedback]{Impact of radiation feedback on the assembly of star clusters in galactic context}

\author[Guillard, Emsellem, Renaud]{
Nicolas Guillard$^{1,2}$\thanks{E-mail: nguillar@eso.org},
Eric Emsellem$^{2,3}$ \&
Florent Renaud$^{4, 5}$
\\
$^{1}$Excellence Cluster Universe, Boltzmannstr. 2, D-85748 Garching, Germany\\
$^{2}$European Southern Observatory, Karl-Schwarzschild-str. 2, D-85748 Garching, Germany\\
$^{3}$Universit\'e de Lyon 1, CRAL, Observatoire de Lyon, 9 av. Charles Andr\'e, F-69230 Saint-Genis Laval; CNRS, UMR 5574; ENS de Lyon, France\\
$^{4}$Department of Physics, University of Surrey, Guildford, GU2 7XH, UK\\
$^{5}$Lund Observatory, Department of Astronomy and Theoretical Physics, Box 43, 221 00, Lund Sweden 
}

\date{Accepted 2018 March 29; Received 2017 October 2}

\pubyear{2017}

\begin{document}
\label{firstpage}
\pagerange{\pageref{firstpage}--\pageref{lastpage}}
\maketitle

\begin{abstract}

Massive star clusters are observed in a broad range of galaxy luminosity and types, and are assumed to form in dense gas-rich environments. Using a parsec-resolution hydrodynamical simulation of an isolated  gas-rich low mass galaxy, we discuss here the non-linear effects of stellar feedback on the properties of star clusters with a focus on the progenitors of nuclear clusters. Our simulation shows two categories of star clusters: those for which feedback expels gas leftovers associated with their formation sites, and those, in a denser environment around which feedback fails at totally clearing the gas. We confirm that radiation feedback (photo-ionization and radiative pressure) plays a more important role than type-II supernovae in destroying dense gas structures, and altering or quenching the subsequent cluster formation. It also disturbs the cluster mass growth, by increasing the internal energy of the gas component to the point when radiation pressure overcomes the cluster gravity. We discuss how these effects may depend on the local properties of the interstellar medium, and also on the details of the subgrid recipes, which can affect the available cluster gas reservoirs, the evolution of potential nuclear clusters progenitors, and the overall galaxy morphology. 

\end{abstract}




\section{Introduction}
\label{intro}
Most of the stars seem to form in clustered environments \citep{Lada2003, Mac2004}. In the Milky-Way, $\sim70$\% of spectral O-type stars are located in young clusters or associations \citep{Gies1987, Parker2007}. In nearby starburst galaxies, young star clusters are also strong UV emitters and the sources of at least 20\% of the UV light \citep{Meurer1995}. Star clusters hence represent key components of star formation in galaxies, play an important role in the formation and evolution of their host, connecting the physics of Interstellar Medium (ISM), star formation and feedback. The formation and evolution of star clusters represent a challenge for models and simulations considering the complex coupling between various spatial scales and physical processes \citep[e.g.][]{Li2016, Naab2016, Niederhofer2016, Chatterjee2017, Bekki2017, Lamers2017}. At the massive end, the build-up of clusters requires high-enough gas densities, which exist in the Local Group, for instance in starbursts (\citealt{Portegies2010}, \citealt{Longmore2014}). Such extreme conditions are also expected to be more frequent at high-redshift when the gas fraction of galaxies was high (50\% and above, see \citealt{Daddi2010}). 

Such gas-rich environments host favorable conditions for the assembly of massive clusters and in particular for the seeds of the most massive clusters, the Nuclear Clusters (NCs). Observed at or near the centre of a wide variety of galaxies of all Hubble types \citep[e.g.][]{Carollo1998, denBrok2014}, NCs are among the densest objects in the Universe, with masses from $10^5\Msun$ to $10^8\Msun$ and typical radius of a few parsecs. They are also characterized by multiple stellar populations spanning ages from 10~Myr to more than 10~Gyr \citep[e.g.][]{Lee1999, Rossa2006, Seth2006, Perets2014}. The main formation scenarios for NCs are nuclear inflows of gas leading to in-situ formation \citep{Milosav2004}, dry-merger of clusters \citep{Tremaine1975} or a combination of the two \citep{guillard2016}. However, it is still unclear how these scenarios relate to the properties of the galactic host.

The formation of NCs is a complex question which relies on the coupling of several physical processes (e.g., star formation, stellar feedback) occurring in environments with extreme physical properties. At the time of their formation (up to 10 Gyr ago, see \citealt{Cole2016} and references therein), gas is presumed to be abundant within the galactic disc, which is favorable to the formation of star clusters \citep[e.g.][]{Arca2015}. \citet{guillard2016} for example showed the importance of gas reservoirs of young clusters in the formation and growth of NCs. Such reservoirs are expected to be significantly perturbed by e.g., star-driven feedback.

Stellar feedback and its effect within galaxies have been extensively studied over the years \citep{Hopkins2014, Hopkins2017, Elbadry2016, Nelson2015, Bournaud2010, MacLachlan2015, Krumholz2014, Raskutti2016, Howard2016, Grisdale2017}. Such studies focused on various spatial scales and physical processes. For example, at parsec (pc) and sub-parsec scale, numerical works have investigated the role of photo-ionization \citep{Dale2012, Walch2012, Tremblin2014, Geen2016} or stellar winds \citep{Wareing2017, Rey2017} in the life of molecular clouds and the star formation within them. The model of feedback implemented by \citet{Nunez2017} in simulations of isolated Milky Way using various physical principles (stellar winds from young massive stars, heating by massive stars within Str\"{o}mgren spheres, and limiting-cooling mechanism based on the recombination time of dense H\textsubscript{II} regions) showed that star formation is more extended (in time and space) when all these physical mechanisms are used simultaneously than when pure thermal supernovae (SNe) feedback is used. \citet{Agertz2013} also showed that pre-SNe feedback (i.e. radiative pressure and stellar winds) is efficient at clearing the gas away from star-forming regions, thus making the subsequent heating from SNe even higher. At kpc-scales, other studies showed that stellar feedback is associated with violent events such molecular outflows \citep{Geach2014, Hayward2017} and helps shaping the gaseous content of galaxies \citep{Agertz2015, Agertz2016}. Since feedback is acting directly or indirectly from sub-parsec to kilo-parsec scales, dedicated hydrodynamical simulations had a hard time both covering the full spatial range and extending over long timescales (Gyr). Moreover, most of these studies are based on conditions that are observed in the Local Universe, and it is thus still unclear how stellar feedback affects the ISM and the forming regions of star clusters in gas-rich discs. With present-day supercomputers, we can actually start to address these issues and the impact of stellar feedback from pc to kpc scales.

Individual feedback mechanisms are expected to play different roles in regulating the assembly of star clusters, and their non-linear interplay makes the matter even more complex to study. The aim of the present paper is to examine their relative contributions and determine how they influence the properties of young star clusters. Addressing these topics would allow us to better understand the direct impact that stellar feedback has on the gas properties, and consequently on that of star clusters when they first form (seeds) and evolve (e.g., growth, merging). The context of this study will be that of an isolated gas-rich galaxy. We choose to focus on a galactic stellar mass of about $10^9\Msun$, as this corresponds to the peak of the fraction of nucleated discs \citep{Pfeffer2014}. With such a setup we extend the study of \citet{guillard2016}, which will serve as a reference. To further understand the role of feedback, we use the same set of feedback recipes as in \citet{guillard2016}, switching off all or part of the feedback components in turn, and comparing the properties of the forming and evolving star clusters. The physics recipes and initial conditions we employ in the present paper are similar to those of the reference simulation. In Section~\ref{num_tech}, we briefly describe the numerical methods. In Section~\ref{sec_results}, we compare the properties of the star cluster population when feedback is active or not. We finally provide a discussion and conclude in Section~\ref{discussion}.



\section{Numerical methods}
\label{num_tech}

In this paper, we present numerical simulations that use initial conditions and prescriptions similar to the ones in \citet{guillard2016}. Hence we only present here a summary, the details being given in \citet{guillard2016}. We conducted hydrodynamical simulations of isolated gas-rich dwarf galaxy with the Adaptive Mesh Refinement (AMR) code \ramses \citep{Teyssier2002}. The code solves the equations of Euler on the AMR grid with a maximum refinement of $3.7 \pc$ in the densest gaseous regions of the ($30 \kpc$)\textsuperscript 3 volume. The least resolved cells span 120pc. The code ensures that the Jeans length is always resolved by at least four cells. A particle-mesh scheme is used to solve the equations of motions, with a softening of $7 \pc$ for the gravitational acceleration of the particles coming from the initial conditions (namely the dark matter and the stars included in the initial conditions) and a minimum of $3.7 \pc$ for the stellar particles formed during the simulations (hereafter stars for simplicity) which corresponds to the local finest refinement of the AMR grid. The simulations have been run on the C2PAP facilities (Excellence Cluster, Garching) for about 1 million CPU-hours on 512 cores.

The isolated gas-rich dwarf galaxy we simulate has a stellar mass of $10^9\Msun$, a gas disc whose gas mass fraction is 70\% of the baryonic mass and a Navarro-Frenk-White \citep{NFW1996} dark matter halo component. The latter has a mass of $10^{11} \Msun$ which follows the scaling relation between DM halos and stellar discs as in \citet[]{Ferrero2012}, a concentration of 16 and a virial radius of 120~kpc. We truncate the halo at a radius of 15~kpc thus focusing on the central regions of the galaxy. Both our stellar and gaseous discs have a radial and vertical exponential profile with a scaling radius of 1 kpc and 1.65 kpc (respectively) and a scale height of 250 pc and 165 pc (respectively).

Our simulations use the same recipes for star formation and stellar feedback which we used in \citet{Renaud2013}. The star formation occurs when the gas reaches a density higher than $100\mathrm{cm^{-3}}$. The gas is then converted into stars with an efficiency of 2\% per free-fall time. These stars have a mass $M_*$ of $130\Msun$. We then model stellar feedback coming from these new-formed stars with 3 processes: photo-ionization which creates H\textsubscript{II} regions, radiative pressure \citep{Renaud2013} and type II supernovae \citep{Dubois2008}. 
In more details, the radius of the H\textsubscript{II} region is 
\begin{equation}
 r_\mathrm{HII} = \left(\frac{3}{4\pi}~\frac{L_*}{n_\mathrm{e}^2 \alpha_\mathrm{r}}\right)^{1/3}
  \label{eq:1}
\end{equation}
where $L_\mathrm{*}$ is the luminosity of the central stellar source, $n_\mathrm{e}$ and $\alpha_\mathrm{r}$ the density of electrons and the recombination rate respectively. Within each of these bubbles, we uniformly set the gas at a temperature of $4\times10^4 \mathrm{K}$, i.e. significantly higher than the surrounding warm ISM. Although this temperature is a few times higher than the typical observed value \citep[e.g.][]{Lopez2011}, we have checked that this difference does not affect our conclusions.
The ionization of the ISM is done as followed: to speed-up the computation, one out of every ten stars radiates and ionizes the surrounding ISM with an energy ten times higher than a single source. Considering that star formation occurs in dense gas regions in a clustered way, this assures that all of these regions contain at least one bubble. Namely, the luminosity follows:
\begin{equation}
 L_\mathrm{*} = L_\mathrm{0}~M_*~\eta_\mathrm{OB} \begin{cases}
    1 & \mathrm{for~t_{ff} < a_* \leq 4~Myr}.\\
    \mathrm{(4 Myr)/a_*} & \mathrm{for~4~Myr < a_* < 10~Myr}. \\
    0 & \mathrm{else}.
  \end{cases} \label{eq:lum}
\end{equation}
where $L_\mathrm{0} = 6.3\times10^{46}~\mathrm{s^{-1}~\msun^{-1}}$, $M_*$ is the mass of the star that spawned, $\eta_\mathrm{OB} = 0.2$ is the stellar mass fraction which explodes into SN and $t_\mathrm{ff}$ is the local free-fall time (see e.g. \citealt{Leitherer1999}). If two H\textsubscript{II} regions overlap each other, the code ensures that the ionized volume is conserved and merges the two bubbles if the separation between the two is smaller than their radii. Finally, the momentum feedback which here is carried by H\textsubscript{II} regions, is injected under the form of velocity kicks and is proportional to $L_\mathrm{*}$ (see \citealt{Renaud2013} for details):
\begin{equation}
 \Delta v = s\frac{L_*~h~\nu}{M_\mathrm{HII}~c}~\Delta t 
  \label{eq:3}
\end{equation}
where h is the Planck constant, $M_\mathrm{HII}$ the gas mass of the bubble, c the speed of light and $\nu$ the frequency of the flux representative of the most energetic part of the spectrum of the source. We consider here the luminosity of the Lymann-$\mathrm{\alpha}$ and set $\nu = 2.45\times10^{15}~\mathrm{s^{-1}}$. s is a dimensionless parameter accounting for the multiple electron scattering through the bubble and the decay of energy between each collisions. We set $s=2.5$ as in \citet{Renaud2013}.

Our stars explode as SNe after 10~Myr. The SNe are modeled as Sedov blasts (see \citealt{Dubois2008}). The initial radius of the ejecta is 10~pc. The total mass removed from each cells affected by the blast wave when the SN explodes is $M_* (1+\eta_\mathrm{OB} + \eta)$ where $\eta$ is the mass-loading factor for the winds. The mass loading factor $\eta$ sets the allocation of the momentum between its mass and velocity terms. SNe inject $10^{51}$ erg of energy in kinetic form and the energy released to the gas by the debris is $E_d = \eta_\mathrm{OB} \frac{M_*}{M_\mathrm{SN}} E_\mathrm{SN}$ where $M_\mathrm{SN}$ and $E_\mathrm{SN}$ are respectively the typical progenitor mass and the energy of an exploding type II supernova (i.e. $10^{51}$ erg). The initial Sedov blast wave propagates at a velocity given by $u_{Sedov} = \frac{\sqrt{2}}{5} \bigg[ f_\mathrm{ek}~\eta_\mathrm{OB} \bigg( \frac{\delta x}{\Delta x}\bigg) ^3 \frac{1}{1+\eta_\mathrm{OB} + \eta}\bigg] ^{1/2} u_\mathrm{SN}$ where $f_\mathrm{ek} = 0.05$, $\delta x^3$ is the volume of the cell where the explosion occurs, $\Delta x$ is the radius of the shock from the center of the explosion and $u_\mathrm{SN}$ is the velocity corresponding to the kinetic energy of one SN explosion. The momentum of the blast wave is then added to that of the gas.

Finally, we use the friend-of-friend algorithm HOP \citep{Eisenstein1998} to detect star clusters. The density thresholds for detection are the same for all simulations and as in \citet{guillard2016}. Namely, a cluster is detected when the peak of the local stellar density exceeds 1.5\mpc with an outer boundary limit of 0.5\mpc to prevent the detection of stars in the field. Two clusters are then merged if the saddle density between them is higher than 1\mpc. 



\section{Stellar feedback and the star clusters population}
\label{sec_results}

In this Section, we examine how the star cluster populations are affected by feedback in our simulations. We thus present the results of test simulations for which a part or all of the feedback modules are turned on or off.




\subsection{The star clusters populations}
\label{sec_SCprop}

We first choose to compare two sets of simulations: one from \citet{guillard2016}, which includes all mentioned feedback recipes (see Sect.~\ref{num_tech}), and another one with the same initial conditions but for which feedback is not active from the start. 

\begin{figure}
  \centering
  \includegraphics{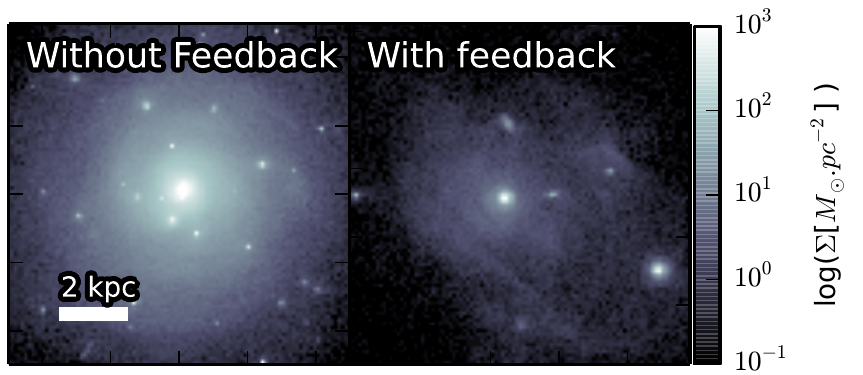}
  \caption{Face-on surface density of the stars formed during the simulations without (left) and with feedback (right). The displayed galaxies have evolved for $940\Myr$. The star cluster population is different between the two cases: without feedback, a massive nucleus forms, surrounded by several tens of smaller and less dense clusters, while with feedback, only 5 clusters orbit around a nuclear cluster.}
  \label{figure:galstars_map}
\end{figure}

Fig~\ref{figure:galstars_map} illustrates the difference in the star cluster populations after $940\Myr$ of evolution between these two simulations. After nearly 1~Gyr of evolution, the star clusters population is already well established in both cases. At these times, the galaxies does not host dense gas clouds anymore, preventing the formation of additional star clusters.

The simulation without feedback exhibits 78 clusters at $t=940\Myr$ with a massive central cluster of $\sim6\times10^8\Msun$, and has formed hundreds of them over that period. Most of these clusters ($\sim 90$\%) formed during the first $50 \Myr$ after the trigger of star formation (at $t=80\Myr$). This strongly contrasts with the outcome of the simulation when feedback is active: only 6 clusters including the nuclear cluster are observed at $t=940\Myr$. The subsequent evolution during the next Gyr is also mild, with no drastic change in the cluster population: 2 clusters are destroyed by cluster-cluster interactions and the nuclear cluster experiences a merger at $t=1.7\Gyr$ (see \citealt{guillard2016} for details). 

\begin{figure}
  \centering
  \includegraphics{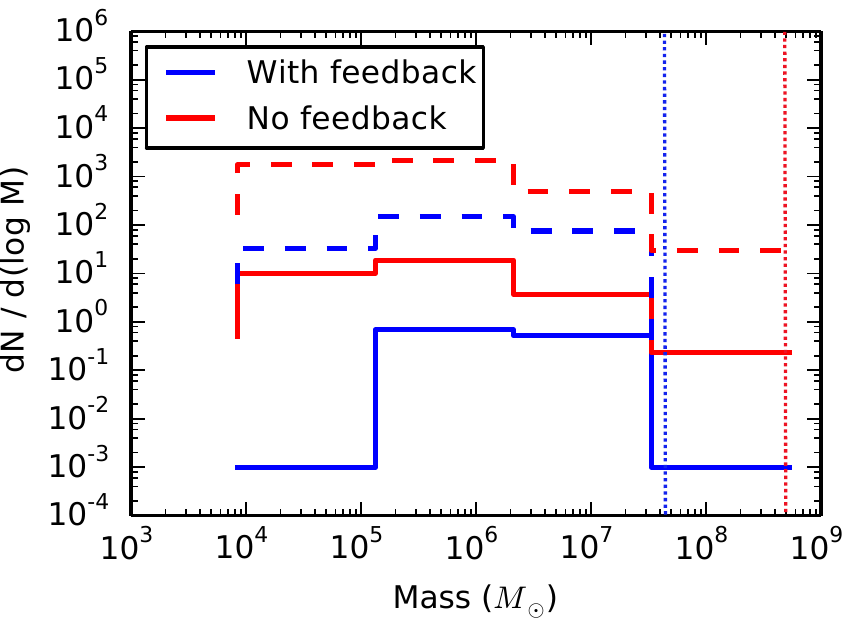}
  \caption{\textit{Solid}: Cluster Mass Function (CMF) for the simulations with (blue) and without (red) feedback at $t=940\Myr$. The dotted lines mark the mass of the nuclei in both simulations. We note that these distributions suffer from low-number statistics. \textit{Dashed}: CMF stacked over $940\Myr$ i.e. for each output between $t=0$ and $t=940\Myr$ (with an average frequency of 1 output every 5 $\Myr$), we calculated the CMF and summed them over that period. They break at $10^4\Msun$ due to the lower limits of the density detection thresholds. We see that at $t=940\Myr$, clusters with mass below $10^5\Msun$ are not observed when feedback is active. Stacked CMF shows that such clusters population existed during the evolution but was either destroyed or grew into more massive clusters.}
  \label{figure:CMF}
\end{figure}

When feedback is active, we notice a lack of star clusters with mass lower than $10^5\Msun$ at $t=940\Myr$ (see Fig~\ref{figure:CMF}): such low-mass clusters are detected at some point (see the stacked distribution of Fig~\ref{figure:CMF}) but either get systematically destroyed by cluster-cluster interactions or merged into more massive clusters, leaving that low-mass bin empty at $t=940\Myr$. It is also worth noting the existence of a population of young stars accumulated in various regions of the disc, but these associations (with stellar densities below our detection threshold) are dispersed by stellar feedback and local variations in the local gravitational potential. 
This contrasts with the simulation without feedback which contains a few tens of such low-mass star clusters. Those clusters are located in the outer regions of the disc and do not interact with one another. This allows the clusters with the lowest mass to survive for more than 2.5~Gyr.   

The nucleus in the no-feedback run is also 10 times more massive ($5\times10^8\Msun$) than that of the reference simulation ($\sim5\times10^7\Msun$). This is a direct consequence of both higher merger rate in the former case (11 mergers for the no-feedback case) and a higher in-situ star formation rate due to the absence of feedback. These mergers supply the nucleus in stars but also in gas which is brought together with the incoming clusters. Such events occur only twice in the reference simulation because of the limited number of clusters. Other surviving clusters of that simulation are gas-free within 10-20 Myr and their mass do not evolve afterwards.

\subsection{The effect on dense gas}
\label{sec:densegas}

\begin{figure}
  \centering
  \includegraphics{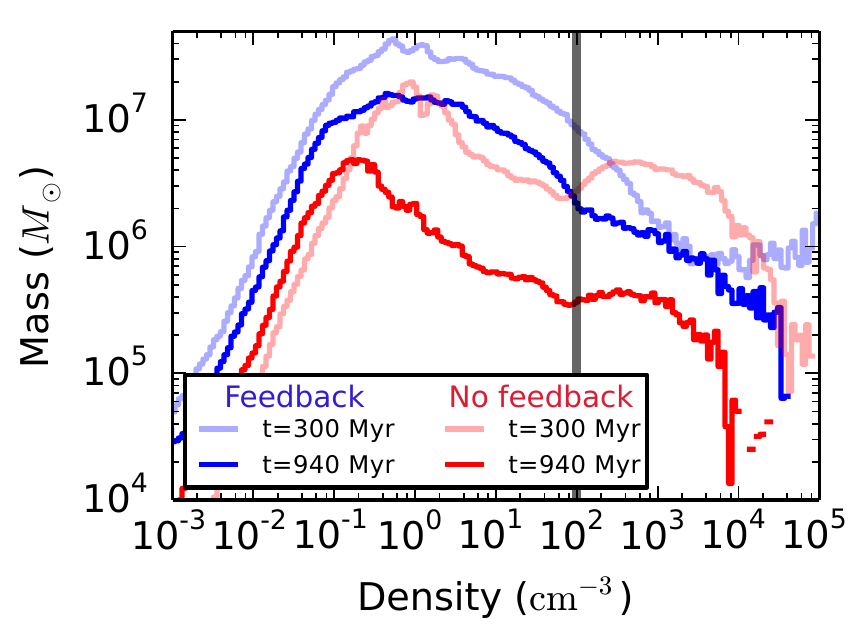}
  \caption{PDF of the gas density for simulations with (blue) and without feedback (red) within the galaxy at various times (showed in the legend). The vertical line marks the star formation threshold. At $t=940\Myr$, less gas is detected in the simulation without feedback than when it is on. This is due to a higher star formation rate (above $1\Msunyr$)}
  \label{figure:pdf_cluster}
\end{figure}

The different formation rate of the clusters between the two cases (15 clusters in 940 Myr with feedback and 320 without feedback) can be interpreted as a direct consequence of the effect of feedback on the availability of dense gas throughout the disc. Fig.~\ref{figure:pdf_cluster} displays the evolution of the Probability Distribution Function (PDF) of the gas density at different times of the life of the two simulated low mass galaxies. 

When feedback is on, the PDF has a log-normal shape for gas densities below $100\cc$ with a peak at $\rho=1\cc$. For higher densities, the shape of the PDF is a power law with a possible mass excess above $3\times10^3\cc$ (e.g., at $t=300\Myr$) corresponding to the central regions of the gas reservoir of the most massive clusters. Such a power-law tail has been interpreted as the convolution of the classical log-normal shape from the turbulent gas with that of the self-gravitating gas clouds \citep{Audit2010, Renaud2013}. As time goes, dense gas is consumed to form stars while part of the lower dense gas cools down and evolves towards higher density, lowering the relative weight of the log-normal part of the PDF. At t= 940~Myr, the nuclear cluster has formed and the power-law seen in Fig.~\ref{figure:pdf_cluster} is associated with its self-gravitating gas reservoir. The excess of mass in the highest density disappears due to the central star formation which consumes dense gas and the following gas dispersion induced by stellar feedback.

The major difference between the simulations with and without feedback occurs around the star formation threshold at $100\cc$. When there is no feedback, a sharp transition is observed at this density, with a lack of gas at lower density and an accumulation of gas with density between $100\cc$ and $300\cc$. A comparison with the simulation with feedback, which displays a smoother transition at $100\cc$, confirms that feedback redistributes dense gas towards lower densities (see e.g. \citet{Grisdale2017} and references therein). This redistribution thus leads to a relative lack of dense gas which can be used to form stars in general, and star clusters in particular. Overall, the gas redistribution induced by the stellar-driven feedback towards low densities is responsible for the smaller number of clusters. This redistribution of dense gas is also responsible for the non-growth mass of star clusters in our simulation with feedback, with the exception of 2 clusters which manage to keep their reservoir (see next Sections). It is also important to note that the absence of stellar feedback does not inhibit the accumulation of dense gas in the disc, since only thermal pressure can oppose the collapse of the clouds. In the next Section, we focus on the processes involved in this redistribution and how they affect the growth of star clusters.

\subsection{Radiative versus Supernova feedback}
\label{sec:SNhydro}

\begin{figure*}
  \centering
  \includegraphics{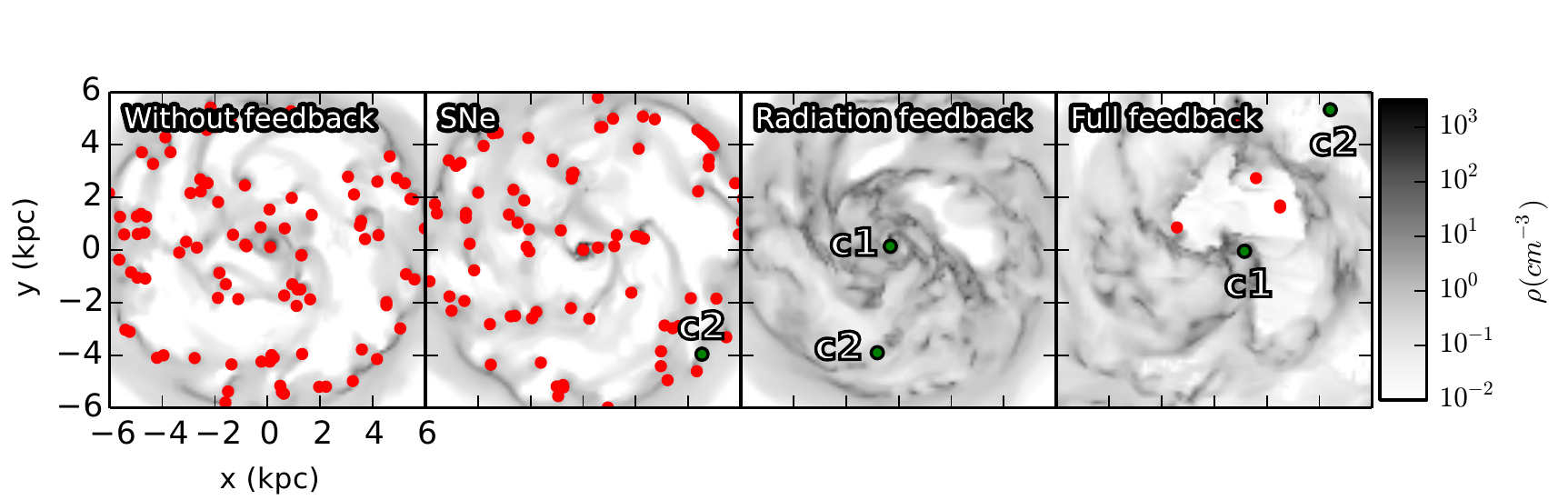}
  \caption{Face-on density maps of the gas at $t=600\Myr$ for simulations using different setups for stellar feedback: without feedback (left), SNe only (middle-left), H\textsubscript{II} regions + radiative pressure (middle-right) and SNe + H\textsubscript{II} regions + radiative pressure (right). The cases with full feedback and without feedback are the same as the ones discussed in the previous sections. The densities are averaged on 1 kpc along the line of sight. The initial conditions are the same for all simulations. Red circles show the position of star clusters. Green circles are the clusters which early evolution will be studied in Fig.~\ref{figure:feed_energy} and Section~\ref{sec:evolution}. Clusters labeled as C1 keep their gas reservoir whereas clusters labeled as C2 expel their gas during the first 10 Myr after its formation. Star cluster populations in the no-feedback and SNe-only cases are similar in terms of number and individual mass. When radiative feedback is on, the ISM becomes less clumpy and the formation of clusters is reduced. The heating from H\textsubscript{II} regions and radiative pressure redistribute the gas towards lower densities and slow down the formation of massive clumps from which star clusters could emerge.}
  \label{figure:map_different_feedb}
\end{figure*}

In the previous Sections, we saw that adding stellar-driven feedback changes, as expected, the properties of the ISM and, consequently, that of the star clusters population. In this section, we further investigate how individual feedback processes impact the growth of star clusters. 

We address this question by running two additional simulations for which we single out SNe feedback, on one hand, and radiative feedback (H\textsubscript{II} regions + radiative pressure), on the other hand. We start with an actual population of star clusters, hence following the mass distribution properties illustrated in Fig~\ref{figure:CMF}, and then focus on the subsequent evolution depending on the implemented feedback schemes. Fig.~\ref{figure:map_different_feedb} illustrates the long-term impact of these recipes by showing the face-on density maps of the gas with the position of all star clusters detected at $t=600\Myr$ (about $500\Myr$ after the trigger of star formation).

Adding only feedback from SNe does not seem to significantly alter the morphology of the ISM on large scales, or the cluster population, compared with the no-feedback simulation. The energy coming from SNe locally increases the temperature of the surrounding gas. Here, SNe are located in dense gaseous regions. Without any mechanisms (e.g. ionization) to disperse such dense gas before the SNe explosions, one could expect the impact of SNe to be less effective on the local environment \citep[e.g.][]{Agertz2013}. Hence, the gas located in dense regions manages to cool down on very short timescales ($\sim 1 \Myr$) leaving most of the dense gas clouds intact. As a result, the seeds for star clusters are not heavily affected by SNe and their populations are similar in both cases (with or without SNe feedback) in terms of mass and numbers. This result was also observed at sub-parsec scales by \citet{Rey2017}, who showed that SNe locally heats the gas, which cools down very rapidly, causing less impact to clouds than stellar winds. We check that the properties of the ISM are not affected by the mass loading factor of the winds (i.e. the amount of gas carried in SNe debris). Using a mass loading of unity, thus affecting a higher gas mass but with smaller velocities, we observe no major differences for the PDF of the gas density or the cluster population in terms of number, size and mass. This suggests that the SNe are not the main actors altering the gas content of cluster-forming regions. We will discuss in Section~\ref{discussion} the impact of the numerical resolution and implementation on this result (see also e.g. \citealt{Smith2017}). 

Major differences arise when radiative feedback is activated: the ISM is less clumpy and more turbulent that in the no-feedback or SNe-feedback cases, with more gas at densities between $1\cc$ and $100\cc$ (of the order of $5\times10^8\Msun$ at $t=600\Myr$ over the entire galaxy). This suggests that the redistribution of the gas towards lower densities (see Section \ref{sec:densegas}) is mainly driven by the radiation from H\textsubscript{II} regions. Hence, this leaves less dense gas over which massive star clusters can form which in turn leads to a lower massive star cluster formation rate: we form only tens of clusters over $500\Myr$ (a few clusters are observed at $t=600\Myr$), as compared again to the hundreds in the simulations without feedback or with SNe only within the same time range. Similar observations can be made with the simulation using all feedback recipes.

\subsection{The ability of clusters to grow}
\label{sec:evolution}

Radiative feedback also affects the growth in mass of star clusters. There are two ways a cluster can gain mass: using a local gas reservoir to convert dense gas into stars or through mergers with other clusters. Since the number of mergers is rather low in simulations using radiative feedback, we focus now on the growth by gas supply. Such process of gas re-accretion occurs in two clusters in our simulations using radiative feedback with the remaining clusters losing this reservoir a few Myr after their formation. Their ability to retain and accrete more of their gas depends on the balance between the gravitational potential of the cluster (i.e. its stellar and gaseous components) and the energy of the gas (internal and injected by feedback). Assuming the systems are in isolation, this balance can be estimated by comparing the total gravitational potential energy of the cluster with the internal energy (which we define here as the sum of the kinetic and thermal energy) of the gas, at a given time (thus ignoring the contribution from e.g., tidal fields). 

\begin{figure*}
  \centering
  \includegraphics{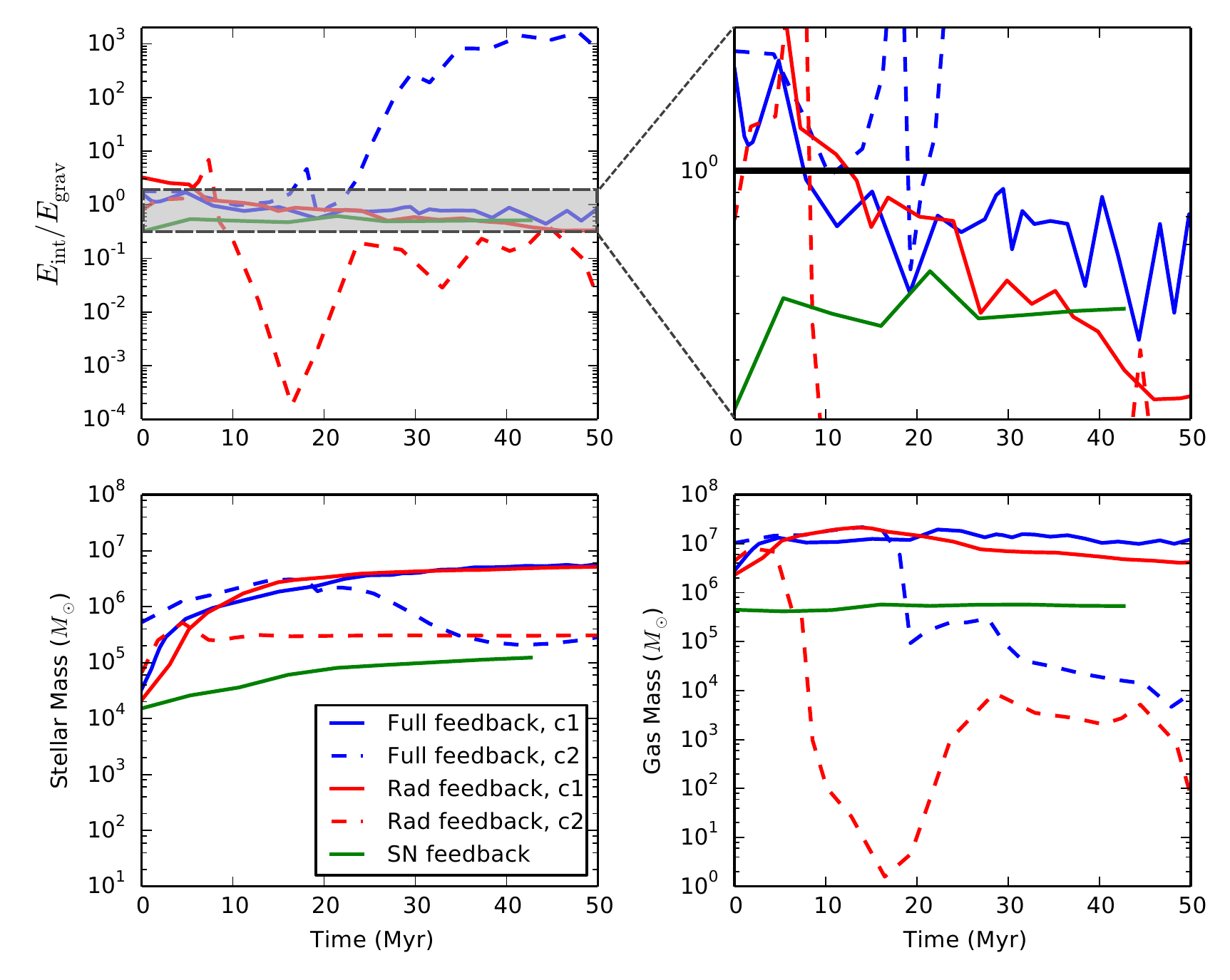}
  \caption{\textit{Top}: Evolution of the ratio of the internal energy of the gas with the gravitational energy of the cluster taken in the 25 pc vicinity of the cluster (the grey area is zoomed in the top-right panel around the unity value for clarity). The time is relative, with $t=0$ being the time of first detection. The different colors represent different simulations. The simulation using only radiative feedback is labeled here as Rad feedback. For the simulations with full and radiative feedback, solid lines display clusters (labeled as C1) which keep their gas reservoir, while dashed lines show clusters (labeled as C2) which expel it.\textit{Bottom left}: Evolution of the stellar mass of the clusters during their first $50\Myr$. \textit{Bottom right}: Evolution of the mass of the gas reservoir.}
  \label{figure:feed_energy}
\end{figure*}

We select a few clusters from the simulations with full, radiative (H\textsubscript{II}+radiative pressure), and SNe feedback, respectively. In
the former two, we choose one cluster which manages to retain its gas for more than 20 Myr (respectively labeled as cluster Full-C1 and Rad-C1, also marked in Fig.~\ref{figure:map_different_feedb}), and one which expels its gas (resp. labeled as Full-C2 and Rad-C2). The C1-clusters are the NC progenitors. The C2 clusters have an initial (i.e. at the time of their first detection) stellar mass density of the order of $4\Msun.\mathrm{pc}^{-3}$ within the inner 25 pc radius, with a slightly higher density for the cluster Full-C2 of $10\Msun.\mathrm{pc}^{-3}$, due to an initially more massive dense gas component in this region. In the simulation with SNe only, we do not observe clusters which expel their gas after their formation. We thus only choose a cluster (SNe-C2) with similar initial stellar density than the other C1 and C2 clusters. 
All these clusters are shown in green in Fig.~\ref{figure:map_different_feedb}.

We follow all these clusters over their first $50 \Myr$. We also limit our estimations of the energies to the inner $25\pc$ around the cluster and systematically check that only one stellar dense structure is included.
We then measure and plot the ratio of the internal energy of the gas to the total gravitational potential energy of the clusters, alongside their stellar and gaseous mass, which we present in Fig.~\ref{figure:feed_energy}. The energies are computed using the mass, velocities associated with the gaseous cells and stellar particles. The kinetic and thermal energies we calculate are defined respectively as $E_\mathrm{kin} = 0.5 \sum_i m_i v_i^2$ and $E_\mathrm{therm} = \frac{1}{M_\mathrm{tot}} \sum_i 3/2 k_\mathrm{B} m_i T_i$ where $m_i$ is the mass of the gas cell \textit{i}-th, $v_i$ its velocity minus the average velocity of the field, $k_\mathrm{B}$ is the Boltzmann constant, $T_i$ its temperature and M\textsubscript{tot} the sum of all $m_i$. The gravitational energy is computed from the gravitational acceleration of the stars and the gas.

In the SNe-feedback simulation, the energy from SNe is immediately dissipated and the gravitational energy dominates (the energy ratio is always lower than unity). Thus, the gas is being retained within the close environment of the cluster, which can then slowly grow its stellar mass. Finally, we note that the gas mass within 25~pc is almost a constant, showing that the consumption of gas by star formation is balanced by the accretion of gas.

When we include the radiation feedback, the C1 clusters retain their gas in both Radiative-only and full feedback cases, despite a bumpier evolution of its total (i.e. stars + gas) energy. During their first 15 Myr, thermal energy associated with H\textsubscript{II} regions is deposited into the ISM. Coupled with the mass of the cluster which is lower than that of the gas by at least one order of magnitude, internal energy dominates the potential with an energy ratio higher than 1. This ratio then slowly decreases in time, reflecting both the steady growth in stellar mass, and the build-up of a massive gas reservoir (the variations of the internal energy are less important than those of the gravitational energy by a factor of 2). The next 35 Myr sees the gravitational energy dominating over the internal energy of the gas, reaching a balance similar to that observed in the SNe-only case. 

The evolution of the total energy of the C2 clusters is clearly different and partly linked to their gas environments. 
After respectively $\sim 5$ and $\sim 20 \Myr$, both the stellar and gas masses of C2 clusters suddenly drop, the latter by several orders of magnitude. This follows the formation of a bubble around the cluster which heats up the gas. During these few Myr, the cluster enters a depleted region of (dense) gas in the disc, reducing the chances for the cluster to accrete more material. This actually leads to a simultaneous decrease of the internal energy of the gas and of the gravitational energy. The relative decrease between these two components determines the outcome energy ratio. For Full-C2, the mass of the gas reservoir significantly decreases after about 18~Myr as the cluster bathes in a hot $5\times10^4$ K ISM. This hot gas is not dense enough to be gravitationally bound to the cluster and the thermal energy dominates, leading to a significant increase of the energy ratio. For HII-C2, as thermal energy from H\textsubscript{II} bubbles is deposited into the ISM, the cluster enters a low density gaseous region of the disc. The gas escapes from the cluster gravitation and the energy ratio dramatically decreases. In both cases, the C2 clusters are almost cleared of their gas reservoirs in a few tens of Myr, and these are not replenished via accretion from the local environment. Note that the decrease of the gravitational energy of these C2 clusters also allows the stars with the highest kinetic velocity to escape.

\begin{figure}
  \centering
  \includegraphics{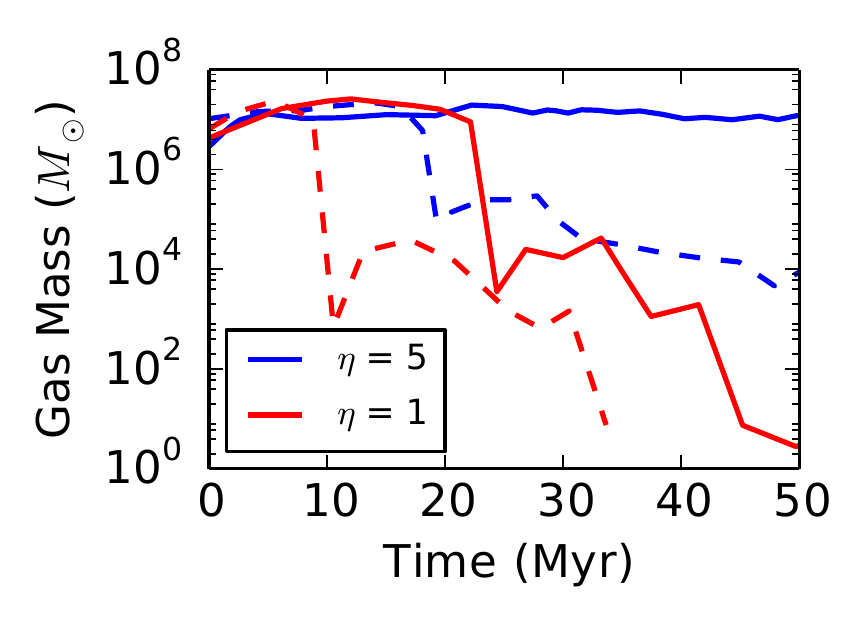}
  \caption{Early evolution of the mass of gas reservoirs in star clusters in simulations using a mass loading factor $\eta$ of 5 and 1 (blue and red respectively). In the latter case, the chosen star clusters have survived up to t=600 Myr and lost their reservoir. Time is relative, with t=0 Myr being the time of formation. Each style shows a different star cluster.}
  \label{figure:mass_loading_factor}
\end{figure}

Similarly to the SNe-only case, we test if the mass loading factor from SNe, now coupled with radiative feedback, affects the early evolution of the mass of star clusters. We thus conduct another simulation with all feedback recipes active and set $\eta = 1$. In such conditions, at $t=600 \Myr$, two clusters of $\sim4\times10^5\Msun$ without gas reservoirs are detected. We also note that no nucleus forms by that time, in contrast with the reference simulation. The early evolution of the gas reservoir of the two detected clusters is shown in Fig.~\ref{figure:mass_loading_factor} and compared with the clusters C1 and C2 of the reference simulation.
When the mass loading factor is 1, the early evolution of both gas reservoirs of star clusters is similar to the cluster C2 but anticipated by 5 or 10 Myr. We note that the first gas-clearing episode (occurring 10 and 21 Myr after the cluster formation) seems to be more efficient than in the reference simulation by an order of magnitude in mass. An intuitive explanation for this is that a low mass loading factor brings a larger volume of gas towards lower densities, which then facilitates the dissipation of the gas reservoir by radiative feedback. This shows the impact of non-linear coupling of different feedback processes (e.g. the radiative feedback and the SNe with the mass loading of the winds) on gas reservoirs and their potential ability to prevent the growth of massive clusters and NC progenitors.  We finally note that, past the first depletion of the reservoir, some amount of gas is brought into the clusters (e.g. small increases of the mass reservoir at t=10 and 29 Myr for the red-dashed cluster). Since the mass loading factor also changes the velocity of the gas carried in the debris, some gas is able to come back onto the cluster. However, this gas is not dense enough to form new stars, and the stellar mass of the cluster does not change drastically after the first expulsion.

Overall, the evolution of star clusters is determined by a fragile balance between their own gravity and the physical properties of their gas. SNe seem to play a role only if they are coupled with radiative feedback. The gas density and temperature of the environment in which clusters evolve are major factors that can change the outcome of the gas reservoir. Massive star clusters are likely to grow if their density allows them to keep dense gas bound to them and if they continuously evolve in a dense gas environment during their first tens of Myr.




\section{Discussion \& Conclusion}
\label{discussion}

Using hydrodynamical simulations of isolated gas-rich galaxies with different radiation feedback setups, we find that :
\begin{itemize}
   \item SNe alone are (mostly) inefficient at affecting the gas reservoir and the early growth of star clusters. Feedback associated with H\textsubscript{II} regions and radiative pressure seem to have a more significant impact, and are thus important components in the early life stages of star clusters. 
   \item When radiation feedback is included, the growth of stellar clusters via gas accretion depends on the ability of the cluster to retain and/or replenish its gas reservoir. That ability is closely tied to the local environment in which the cluster goes through during the first tens of Myr, and to the corresponding availability of dense gas around the cluster as it orbits within the disc. This allows the galaxy to develop two categories of star clusters population: those from which feedback expels the gas reservoir shortly after their formation, and those in a denser environment around which feedback fails at totally clearing the gas. We also note that low mass loading factors (i.e. 1 in our case) for the SN blast coupled with radiative feedback can efficiently disperse dense gas, thus preventing the growth of star clusters.
   \item In HII or Full feedback simulations, we would expect only the massive end of the cluster distribution to survive (with the typical mass of a few $10^5$ to a few $10^6 \Myr$ in the present case) depending on the specific locations/trajectories of the clusters.
\end{itemize}

These conclusions align with several studies. Based on timescale estimations, \citet{Krumholz2009} already argued that SNe should play a limited role as a source of feedback in star clusters since H\textsubscript{II} regions inject their energy immediately after the star formation and do not have delays like SNe. In addition, results from \citet{Li2017} suggest that the impact of SNe is weaker in high-density environments which is where our clusters form (see their Figure 10). On galactic scales, \citet{Butler2017} showed that the combination of $\mathrm{H_2}$ dissociation, photo-ionization from extreme ultraviolet photons and SNe leads to different properties of the gas in terms of temperature and different spatial distribution of young stars compared to a case where only SNe are active. Our experiments point towards the same trends, emphasizing the paramount role of non-linear multi-component feedback, in particular in the formation of massive stellar objects.

The impact of feedback in numerical simulations obviously depends on the employed subgrid implementations of SNe and radiative feedback. Our work suggests that SNe alone are inefficient at disturbing the gas properties and the production of stars (see also \citealt{Smith2017}). Similar results have been observed when thermal feedback is used: it has been suggested that such inefficiency may be due to the fact that the SNe energy is distributed over too much mass, meaning that the temperature of the heated ISM around the SN is too low \citep{Dalla2008}. A potential measure to stop the gas from over-cooling with only SNe would be to use a mechanical feedback like in \citet{Smith2017}, as it injects momentum depending on the relevant scale of the SN remnant. In their work,  \citet{Smith2017} showed this technique has the advantage to reach numerical convergence of the star formation rate even at resolution of 8.1 pc. Nonetheless, it is unclear how the non-linear coupling between this kind of feedback and radiation will affect the properties of the ISM and those of star clusters.

\begin{figure}
  \centering
  \includegraphics{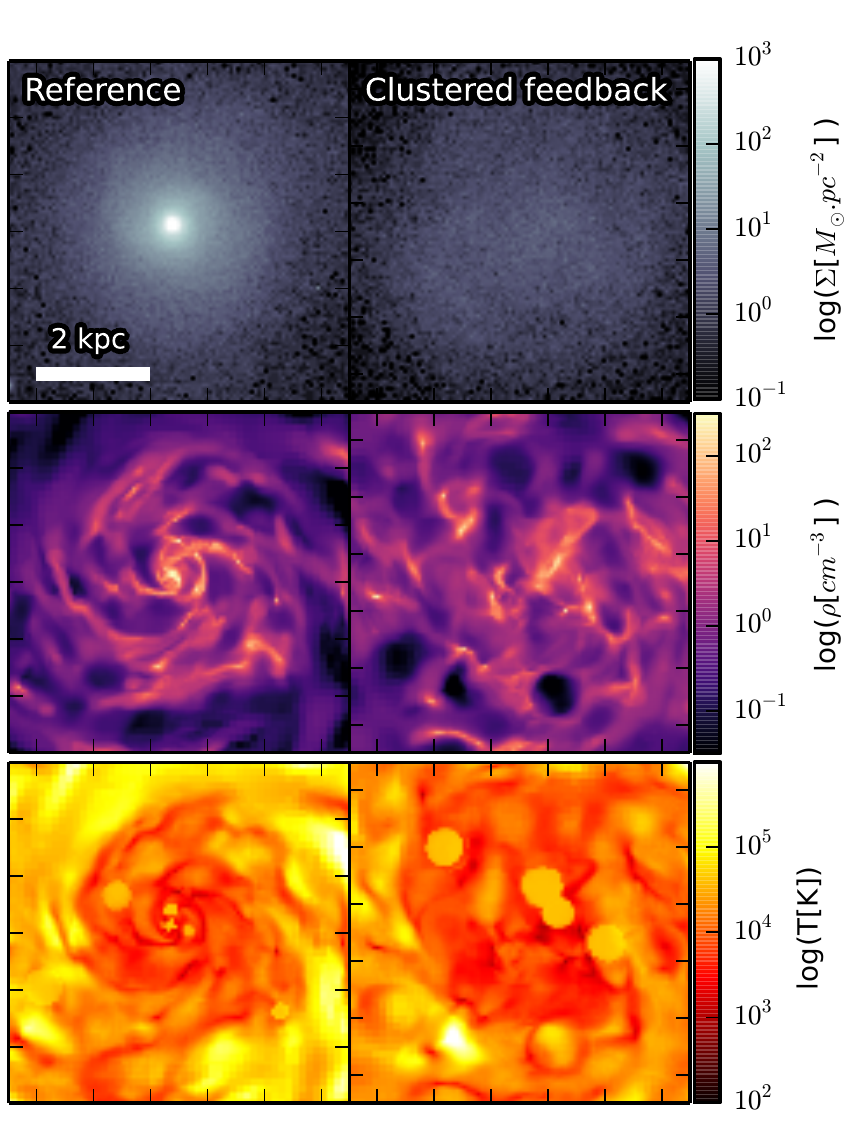}
  \caption{Stellar (\textit{top}) and gas (\textit{middle}) densities and gas temperature (\textit{bottom}) maps using 2 different feedback setups at $t=2.5\Gyr$. The left panels show our reference simulation. On the right panels, we generate 10 times fewer H\textsubscript{II} regions but each of these are 10 times more energetic individually (to conserve the total energy injected into the ISM of the whole disc). This generates differences in the ISM and star clusters properties which modify in the end the morphology of our dwarf galaxy.}
  \label{figure:map_beta}
\end{figure}
The implementation of the ingredients used for our radiative feedback also plays a role since the properties (size, mass, energy, etc) of H\textsubscript{II} regions may vary, depending on the local conditions. In a gas-rich environment, such variations could directly impact the star cluster populations and its evolution. To illustrate this point, we can artificially increase the energy input for individual H\textsubscript{II} regions while conserving the total injected energy (by lowering the number of H\textsubscript{II} regions), the radial extent of the associated bubbles increasing accordingly (by a factor of about 10). Because the radius of the bubbles increases, we refer to this setting as clustered feedback.

Figure~\ref{figure:map_beta} illustrates how such an imposed change in the energy injection scheme naturally perturbs the morphology of the galaxy by, ultimately, preventing the formation of a nucleus. Indeed, the local heating by larger H\textsubscript{II} regions induces a decrease in the gas local density which alter its properties on a larger scale (i.e. kpc scales) and with it, the location and number of the forming star cluster sites. Larger gas-rich volumes are heated, and a larger fraction of dense gas is being shifted to lower densities, hence compromising the further growth of potential NC progenitors. Hence, calibrating the radiative feedback is of crucial importance if one wants to study the properties of gas in discs and the morphology of galaxies in general. Also, as the effect of radiative feedback depends on the gas density in the disc and thus on the spatial resolution, such calibration should be different between simulations of isolated discs and cosmological simulations.

Stellar feedback further encompasses several more coupled processes other than SNe or radiative feedback, some of which are not included in the present simulations (e.g., photoelectric heating, cosmic rays). Discussions on the relative effects of these feedback processes can be found (for dwarf galaxies) in e.g., \citet{Kima2013, Kimb2013, Hu2017,Forbes2016}. Feedback from low and intermediate-mass stars could also impose a time delay of the star formation \citep{Offner2009,Dale2017}, thus potentially lowering the number of clusters: such an effect has been ignored in our simulation as we do not sample the initial mass function. Magnetic fields, which could suppress the expansion of H\textsubscript{II} regions \citep{Krumholz2007, Peters2011} are also not taken into account: their inclusion might enhance the ability for a cluster to grow since less volume would be heated by the bubbles. Finally, cosmic rays are currently thought to be generated by supernovae and massive stars winds which are conditions encountered in regions of massive star formation \citep{Veritas2009, Bykov2017}. The heating of the gas from cosmic rays might be an obstacle to a high-dense gas reservoir and thus to the growth of NC seeds.

In our simulations, these seeds all reach a mass of $\sim 10^6\Msun$ after a few Myr. High-resolution studies of individual Giant Molecular Clouds, such as \citet[]{Dale2012} showed that clusters above $\sim 10^6\Msun$ have a high enough escape velocity to prevent H\textsubscript{II} regions to efficiently remove the gas from the clusters. This further illustrates the potential ability of the young massive clusters in general and NC progenitors in particular to retain their gas reservoirs in dense environments, like gas-rich galaxies and mergers. On the other hand, the lower mass clusters ($\lesssim10^5 \Msun$) are strongly impacted by ionizing feedback.
We also note that the mass range and the parent GMC of our C2 clusters are in agreement with recent work from \citet{Howard2017} which studied the impact of the inclusion of radiation feedback on the efficiency of cluster formation, and established a relation between the maximum mass of a star cluster and the mass of the parent cloud ($\mathrm{M_{cluster,max} \propto M_{cloud}^{0.81}} $).
 
Most of the clusters we have studied have densities of $\sim 10\Msun.\pc^{-3}$. Our spatial and mass resolution does not allow us to form low mass bound systems such as associations and open clusters. The typical stellar density for these objects goes from 0.01 to 1~Star.pc$^{-3}$ for associations and open clusters, respectively. This would require sub-parsec resolution which is beyond the scope of this study. Lower mass clusters (e.g., $10^3 - 10^4 \Msun$) are expected to be more vulnerable to feedback disturbances in a gas rich environment. Associations generally disperse over timescales of $10\Myr$ and we would expect feedback to participate in the dissolution process. 

Galactic and extra-galactic environments are also likely to affect the properties of star clusters such as their mass or density. Open clusters are mostly observed in spiral arms \citep{Dias2002} whereas more massive (forming) clusters are observed in e.g., starbursts and mergers \citep[e.g.,][]{Portegies2010} or central regions \citep{Boker2002}). For the specific cases of NCs, hosted at the centre of galactic discs, some studies \citep{Emsellem2015, Torrey2016} suggest an interplay between star formation and feedback processes, leading to gas accretion-ejection cycles and possibly to complex integrated star formation histories \citep{Feldmeier2015}. In galaxy mergers such as the Antennae, young massive star clusters generate superbubbles of hundreds of parsecs \citep{Camps2017} in the nuclear regions which might take the least dense gas away and hence halt the star formation within these clusters. 

Overall, this work emphasizes the importance of the calibration of feedback recipes, its impact on properties of the ISM and star clusters. We also underline the relevance of a more realistic galactic-scale environment (interactions, gas accretion) for the early formation and evolution of massive clusters.

\section*{Acknowledgements}

We thank the anonymous referee for her/his helpful comments and suggestions that helped improving the paper.
We also thank Oscar Agertz for interesting discussions.
This research was supported by the DFG cluster of excellence 'Origin and Structure of the Universe' (www.universe-cluster.de).
We acknowledge the support by the DFG Cluster of Excellence "Origin and Structure of the Universe". 
The simulations have been carried out on the computing facilities of the Computational Centre for Particle and Astrophysics (C2PAP). 
FR acknowledges support from the European Research Council through grant ERC-StG-335936 and the Knut and Alice Wallenberg Fondation.



\bibliographystyle{mnras}
\bibliography{paper_guillard_feedback_SC} 
\nocite{Georgiev2016} 








\bsp	
\label{lastpage}
\end{document}